\documentstyle[twocolumn,psfig,rotating]{mn}
\def\dd{{\rm d}}
\def\ln{{\rm ln}}
\def\pa{\partial}
\def\grad{\nabla}
\def\pmb#1{\setbox0=\hbox{#1}%
\kern-.025em\copy0\kern-\wd0
\kern.05em\copy0\kern-\wd0
\kern-.025em\raise.0433em\box0}
\def\vV{\pmb{$V$}}

\def\vg{\pmb{$g$}}
\def\vE{\pmb{$E$}}
\def\tvg{\pmb{${\tilde {\vg}}$}}
\def\vnabla{\pmb{$\nabla$}}

\begin{document}
\title[Diffusion in clusters ]
{Gravitational diffusion in the intra-cluster medium}
\author[Chuzhoy \& Nusser ]{Leonid Chuzhoy
and
Adi Nusser\\
Physics Department and The Space Research Institute,
Technion, Haifa 32000, Israel\\
E-mail: cleonid@tx.technion.ac.il,  adi@physics.technion.ac.il
}

\maketitle

\begin{abstract}
  We revisit the process of gravitational sedimentation of helium and
  heavy elements in the intra-cluster medium.  We find that helium
  applies an inward drag force on heavy elements, boosting their
  sedimentation speed to nearly half its own.  This speed is almost
  independent of the mass and the electric charge of heavy elements.
  In the absence of  small-scale magnetic fields, helium sedimentation can
  increase the He/H abundance ratio in the cores of hot clusters by
  three orders of magnitude.  It also steepens the baryonic density
  profile yielding a higher X-ray luminosity, which offers an
  explanation of the observed luminosity-temperature relation.  

If the
  primordial He/H ratio is assumed, then the gas density inferred from
  the observed X-ray emissivity might be underestimated by $ 30\%$ in
  the cores of clusters and overestimated by $7\%$ in the outer
  regions.  The dark matter density on the other hand might be
  overestimated by a factor of 8/3 in the cores and underestimated by
  $ 18 \%$ in outer regions.

\end{abstract}


\section {Introduction}

In the equilibrium state of a multi-component plasma the number
density, $n_{\rm i}$, of particles of mass $m_{\rm i}$, is $n_{\rm
i}\propto {\rm e}^{-m_{\rm i}\phi(r)/kT}$, where $\phi(r)$ is the
gravitational potential.  In the high temperature intra-cluster medium
(ICM), light elements, helium (He) and hydrogen (H), can diffuse fast
enough to reach their equilibrium distribution in a Hubble time.
Larger frictional drag forces work on heavier elements, but still we
expect, at least partial segregation as a result of diffusion.

Diffusion can have important  consequences.
Fabian \& Pringle (1977) suggested diffusion as a possible
explanation for the observed gradients in the iron abundance inside galaxy
clusters.  In their calculations iron ions sediment by 
 a distance comparable to the radius of the cluster within a Hubble time. 
For other elements they predicted  the diffusion
velocity to be proportional to $AZ^{-2}$, where $A$ is the atomic number
and $e Z$ is the charge of the ion.  These calculations assumed 
that only  protons apply appreciable drag forces on 
heavy elements.  Rephaeli (1978)
claimed that by  neglecting helium drag  
 Fabian \& Pringle overestimated the iron diffusion rate.

Helium sedimentation can potentially change the global observational
properties of X-ray emission from rich clusters. As has been noted by
Qin \& Wu (2000) current estimates of cluster masses from X-ray
observations, which rely on the assumption of constant mean molecular
weight, can be off by $\sim 18 \%$ if helium is concentrated inside
the core.  Here we also point out that by steepening the baryonic
density profile helium sedimentation increases the X-ray emissivity in
hot clusters.

In this paper we revisit the calculation of element sedimentation in
the ICM.  We confirm the claim by Rephaeli that helium drag on iron is
comparable to that of protons. However, instead of hindering the
sedimentation of iron and other heavy elements we show that helium
acts as a catalyst. Our analysis includes electric fields that are
inevitably generated by segregation of charged elements.  Although
these fields reduce the diffusion rate,  the sedimentation
time-scales of heavy elements  are still several times
shorter than previous estimates.

The paper is organized as follows. In \S 2 we present the basic equations
and estimate the sedimentation speeds and time-scales. In \S 3 we
discuss the equilibrium distribution as a limiting case of 
element sedimentation. We conclude in \S 4.

\section{The equations of element diffusion}

We write the equations governing the evolution of individual species
in an ICM of any composition. We do not consider magnetic fields in
this paper, but include electric fields which must exist in any
ionized plasma in a gravitational field (Eddington 1926).  Let the
ICM be made of any number of species each made of particles
characterized by mass $m_{\rm i}$ and electric charge $q_{\rm i}=e
Z_{\rm i}$. We denote the local number density and velocity of a patch
of matter of each species by $n_{\rm i}$ and $\vV_{\rm i}$. Then, the mass
density is $\rho_{\rm i}=n_{\rm i} m_{\rm i}$ and the pressure is
$P_{\rm i}=n_{\rm i} kT$, where we have assumed that the ICM is in
local thermal equilibrium so all species share the same temperature
$T$. With this notation the equations are 
\begin{eqnarray}
\label{h2}
\frac{\pa \vV_{\rm i}}{\pa t} +\vV_{\rm i}\cdot \vnabla \vV_{\rm i}
=-\frac{\vnabla P_{\rm i}}{\rho_{\rm i}}+\vg+\frac{q_{\rm
i}\vE}{m_{\rm i}}+\sum_{j}(\vV_{\rm i}-\vV_{\rm j})/\tau_{\rm ij} \, ,
\end{eqnarray}
where $\vE$ is the electric field, and $\vg$ is the gravitational
field (force per unit mass). The constants $\tau_{\rm ij}$ are the
time-scales for the drag forces acting on the species $i$ as a result
of Coulomb interactions with species $j$.  Momentum conservation
implies that $\tau_{\rm ij}=(\rho_{\rm j} /\rho_{\rm i})\tau_{\rm
ji}$.  If $m_{\rm i}\gg m_{\rm j}$ then $\tau_{\rm ij}$ can be
approximated by (Spitzer 1968)
\begin{eqnarray}
\label{Sp}
\tau_{\rm ij}&=&\frac{3(2\pi)^{1/2}(kT)^{3/2}}{8\pi Z_{\rm i}^2Z_{\rm
j}^2e^4n_{\rm j} \ln \Lambda}\frac{m_{\rm i}}{m_{\rm j}^{1/2}}\\& =& \nonumber
9.4\times 10^6\left(\frac{T}{10^8 {\rm K}}\right)^{3/2}
\left(\frac{{\rm \ln}\Lambda}{40}\right)^{-1} \left(\frac{n_{\rm j}}{10^3
{\rm m}^{-3}}\right)^{-1}\\&\nonumber\times&\left(\frac{m_{\rm i}}{m_{\rm
p}}\right) \left(\frac{m_{\rm j} }{m_{\rm
p}}\right)^{-1/2}\left(Z_{\rm i} Z_{\rm j}\right)^{-2}\, {\rm Yr}\, ,
\end{eqnarray}
where 
the ${\rm ln}
\Lambda$ is the Coulomb logarithm.
The equations (\ref{h2}) must be  supplied by 
additional relations that specify  the electric field. 
To create an electric force comparable with gravitational and pressure forces 
a tiny fractional  charge excess (${Gm_{\rm p}^2}/{e^2}\sim 10^{-36}$) 
is sufficient.
The additional relations can then be obtained by 
assuming charge neutrality and zero  electric currents, i.e.,
\begin{equation}
\sum_{i} n_{\rm i} q_{\rm i}=0 \, ,\quad {\rm and} \quad \sum_{\rm i}
n_{\rm i} q_{\rm i}\vV_{\rm i}=0\, .
\label{h5}
\end{equation}

\subsection{Sedimentation speeds and time-scales}

Multiplying the equations (\ref{h2}) by $n_{\rm i} m_{\rm i}/\sum n_{\rm
i} m_{\rm i} $ and summing over all species yields
\begin{eqnarray}
\frac{\vnabla P}{\rho}= \tvg \, ,
\label{h1}
\end{eqnarray}
where $\rho=\sum n_{\rm i} m_{\rm i}$ is the total local density of
all species, $\vV=\sum \rho_i \vV_{\rm i} /\rho$ and $P=\sum 
P_{\rm i}$
are, respectively, the mass-weighted mean velocity and total pressure, and
$\tvg=\vg-\pa \vV/\pa t - \vV \cdot \vnabla \vV$ is the gravitational
field in the frame of reference moving with the velocity $\vV$.  The
terms involving the electric and drag forces have disappeared by
virtue of charge neutrality and momentum conservation, respectively.
Since $|\vV_{\rm i}-\vV|$ is much smaller than the thermal velocities,
the velocity dispersion
term $\sum \rho_i (\vV_{\rm i}-\vV)\vnabla (\vV_{\rm i}-\vV)/\rho$
is negligible compared to the pressure and gravity terms 
 we have not included it in the equation (\ref{h1}).

At the initial stage of the diffusion process the dominant light
species have the same
distribution, which gives ${\grad (n_{\rm i}
kT)}/{(n_{\rm i} m_{\rm i})}=(\mu m_{\rm p}/m_{\rm i})\tvg$, where
$\mu=(1/m_{\rm p})
\sum n_{\rm i} m_{\rm i}/\sum n_{\rm i}$ is the mean molecular
weight.
Thus equations (\ref{h2}) yield
\begin{eqnarray}
\label{h3}
(\mu m_{\rm p}/m_{\rm i}-1)\tvg+\frac{q_{\rm i} \vE}{m_{\rm
i}}+\sum_{j}(V_{\rm i}-V_{\rm j})/\tau_{\rm ij }=0\; .
\end{eqnarray}
Using equations (\ref{h5}) and (\ref{h3}) we find that the electric
field is $\vE=\mu \tvg/e$. 
From (\ref{Sp}) $\tau_{\rm
ij}\propto n_{\rm j}^{-1}m^{-1/2}_{\rm j}$, so because of  the  low
abundance of heavy ions and the low mass of electrons,
only the drag terms from  protons
and helium ions is important. So we are left with a single equation
for each element
\begin{eqnarray}
\label{hs}
 \left(\frac{1+Z_{\rm i}}{A_{\rm i}}\; \mu-1\right)\tvg+
\frac{\vV_{\rm p}-\vV_{\rm i}}{\tau_{\rm ip}}+\frac{\vV_{\rm \alpha}-\vV_{\rm i}}{\tau_{\rm i\alpha}}=0 \; .
\end{eqnarray}

We can now estimate the velocity of each species relative to $\vV_{\rm
p}$, the velocity of the proton fluid at each point. The advantage of
using velocities relative to protons is the independence of the
results from physical processes other than diffusion (radiative
cooling, heating by supernovae etc). Even though these processes can have
important dynamical effects on the cluster, they do not affect the
expressions we develop for the relative velocities and element
abundance.

From (\ref{hs}) the helium velocity is 
\begin{eqnarray}
\vV_\alpha-\vV_{\rm p}
=(3\mu/4-1)\tvg\; \tau_{\rm \alpha p}\approx -0.56\tvg\; \tau_{\rm \alpha p}.
\end{eqnarray}

The relative sedimentation speed of heavy elements, $\vV_{\rm
i}-\vV_{\rm p}$, can easily be related to that of helium
$\vV_\alpha-\vV_{\rm p}$.  Heavy elements experience an upward proton
drag and inward helium drag forces. According to (\ref{Sp}),
$\tau_{\rm ij}\propto Z_{\rm i}^{-2}$ and for large $Z_{\rm i}$ these
drag terms dominate all others in (\ref{hs}).  Thus we are left with
\begin{eqnarray}
\label{vel}
(\vV_p-\vV_{\rm i})/\tau_{\rm ip}\approx (\vV_{\rm i}-\vV_{\alpha})/\tau_{\rm i\alpha}.
\end{eqnarray}
Substituting $\tau_{\rm i\alpha}$ and $\tau_{\rm i p}$ from (\ref{Sp}) in 
the last equation yields
\begin{eqnarray}
\frac{\vV_{\rm i}-\vV_{\rm p}}{\vV_{\alpha}-\vV_{\rm p}}\approx
\left(1+\frac{n_{\rm p}}{8n_\alpha }\right)^{-1}= 0.4\, ,
\end{eqnarray}
where we have taken $n_\alpha/n_{\rm p}=0.08$.
Thus, contrary to the prediction of Fabian \& Pringle (1977), who by
neglecting helium drag obtained $(V_{{\rm p}}-V_{\rm i})\propto A_{\rm i}
Z_{\rm i}^{-2}$, the diffusion speeds are approximately the same for
all heavy elements.
 Including all forces in the calculation increases slightly this
result (by a factor of $\sim 1+4/Z_{\rm i}$)
\begin{eqnarray}
 \frac{\vV_{\rm i}-\vV_{\rm p}}{\vV_{\alpha}-\vV_{\rm p}}=0.4 
\left[1+\frac{3A_{\rm
i}}{Z_{\rm i}^2}-1.8\left(\frac{1}{Z_{\rm i}}+\frac{1}{Z_{\rm
i}^2}\right)\right].
\label{ratspeed}
\end{eqnarray}
There is also  a correction of a similar magnitude  if
 the initial distribution of heavy elements is different from that of 
 light  elements.

 The relation (\ref{ratspeed}) is independent of the physical state of
 the ICM.  To obtain the relative velocities we have to know the
 temperature, the density and the gravitational acceleration.  Taking
 in (\ref{Sp}) ${\rm ln}
\Lambda\approx 40$ as the typical value for the ICM,
we find the helium velocity relative to protons to be
\begin{equation}
 \vV_{\alpha}-\vV_{\rm p}=5\times 10^4 \tvg_{_{-9.5}}{n_{\rm p}^{-1}}_{_{-3}}
T_{_8} \hspace{3mm} {\rm m s^{-1}},
\label{speeds}
\end{equation}
where $\tvg_{_{-9.5}}=\tvg/(10^{-9.5} {\rm ms^{-2}})$, ${n_{\rm
p}}_{_{-3}} =n_{\rm p}/(10^3 {\rm m^{-3}})$, and $T_{_8}=T/10^8$ K.
This is lower by $\sim 40\%$ than the result of Qin \& Wu (2000), the
main difference is the inclusion of electric field in our calculation.
As seen from equation (\ref{speeds}) the diffusion speeds depend on
the local density and temperature. However, a single diffusion
time-scale is obtained if the gas and dark matter both follow an
isothermal spherical density profile $\rho(R)\propto R^{-2}$ and the
gas is in hydrostatic equilibrium ($\tilde g=g$) with constant $T$.
In this case $V_{\rm i}-V_{\rm j}\propto g/n_{\rm p}\propto R$, which
together with continuity equation, $\dd n_{\rm i}/\dd t=-R^{-2}
\partial_R(R^2 V_{\rm i})$, yields,
\begin{eqnarray}
\frac{\rm d}{{\rm d}t}\ln\left(\frac{n_{\rm i}}{n_{\rm j}}\right)=
\frac{3(V_{\rm j}-V_{\rm i})}{R} \; .
\label{cont}
\end{eqnarray}
This motivates us to  
define the time-scale for diffusion between two species as 
\begin{equation}
\tau_D=\frac{R}{3(V_{\rm i} -V_{\rm j} )},
\end{equation}
which for helium relative to  protons gives
\begin{eqnarray} 
\tau_D= 3\times 10^9 \left(\frac{f_{\rm b}}{0.1}\right)
\left(\frac{T}{10^8 {\rm K}}\right)^{3/2} {\rm Yr}\; ,
\label{heliumtau}
\end{eqnarray}
where $f_{\rm b}$ is the baryonic mass  fraction in the cluster.

\section{The Equilibrium Distribution}

The diffusion time-scale of helium as seen from (\ref{heliumtau}) is
comparable to the Hubble time. It is thus prudent to examine in detail
the   equilibrium distribution of hydrogen and helium, which is the final
product of diffusion.

We assume a spherically symmetric  cluster in which the dark mass
 inside a radius $r$  is given by (Navarro, Frenk \& White 1997)
\begin{eqnarray}
M_{_{\rm dm}}(r)=4\pi \rho_s r_s^3\left[
\ln \left(1+r/r_s\right)+\frac{1}{1+r/r_s}
\right]
 \, ,
\label{nfw}
\end{eqnarray}
where $\rho_s$ and $r_s$ are  constants.
 The 
gas density profile, $\rho$, is determined by 
 the equation of hydrostatic equilibrium (see eq.\ref{h1}),
\begin{eqnarray}
\frac{kT}{\mu m_{\rm p}}\frac{\dd}{\dd r} \ln
\left(\frac{\rho}{\mu}\right)=
-\frac{GM_{_{\rm dm}}(r)}{r^2} \; ,
\label{hstat}
\end{eqnarray}
where 
  we have neglected the contribution of baryons
to gravity and assumed constant  
temperature, $T$, throughout the ICM.
In the absence of  sedimentation  $\mu$ is constant and 
 the solution to (\ref{hstat}) is (Makino, Sasati, \& Suto 1998)
\begin{eqnarray}
\rho(r)=\rho(0)e^{-\mu \eta}(1+r/r_s)^{\mu\eta r_s/r}\; ,
\label{nodiff}
\end{eqnarray}
where $\eta=4\pi G m_{\rm p} \rho_s r_s^2/kT$.
Sedimentation  introduces a dependence of $\mu$ on $r$ 
and the above analytical
solution is no longer valid. The abundances of the 
various elements are  then
 determined by the equation of  hydrostatic equilibrium for each
element separately and  the condition for local charge neutrality,
\begin{eqnarray}
\label{difeq}
\frac{kT}{m_{\rm i}}\frac{d(\ln n_{\rm i})}{
dr}&=&-\frac{GM_{_{\rm dm}}(r)}{r^2}-\frac{q_{\rm i}E(r)}{m_{\rm i}},\\ \sum_{\rm i}
n_{\rm i}(r) q_{\rm i} &=&0\; .
\end{eqnarray}
Taking $m_\alpha=4m_{\rm p}$ and neglecting the electron mass, these equations
have the solution
\begin{eqnarray}
n_{\rm p}&=&6 C_1\left[f(r)+f^{-1}(r)-1\right]/ h^2(r)\; , \\
n_\alpha&=&C_1\left[f(r)+f^{-1}(r)-1\right]^2/ h^2(r)\; ,
\label{analytic:sol}
\end{eqnarray}
where $f(r)=\left[C_2h^5(r)-1+\sqrt{(C_2h^5(r)-1)^2-1}\right]^{1/3}$,
$h(r)=(1+r/r_{\rm s})^{\eta r_{\rm s}/r}$, 
 and   $C_1$
and $C_2$ constants.  The behavior of the
solution in the inner and outer regions is easily understood as
follows.  In the inner regions helium is dominant, thus $\mu=4/3$ and
$E=\mu g/e=4g/3e$. From this follows $n_p\propto (1+r/r_s)^{-\eta/3}$
and $n_\alpha\propto (1+r/r_s)^{4\eta/3}$.
 Note that, since the total force felt by protons
becomes repulsive ($eE>mg$), their density is falling towards the
center.  The outer regions consist almost entirely of hydrogen plasma,
thus $\mu=1/2$ and $E=\mu g/e=g/2e$. This gives $n_p\propto
(1+r/r_s)^{\eta/2}$ and $n_\alpha\propto (1+r/r_s)^{3\eta}$, and so
$n_\alpha\propto n_p^6$, in agreement with Gilfanov \& Syunyaev (1984).

\begin{figure} 
\centering
\mbox{\psfig{figure=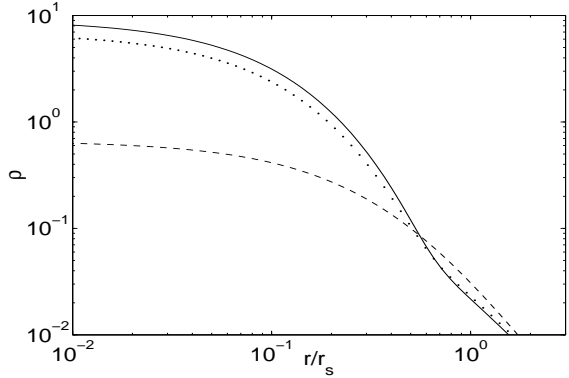,height=2.0in,width=3.0in}} \caption{
Gas density profiles.  The solid and dashed lines correspond
to equilibria with and without diffusion,
respectively. The dotted line corresponds to the profile inferred from
the X-ray emissivity by assuming constant abundances in the case
with diffusion. The density is normalized so that the total baryonic mass 
within  $3r_s$ is unity.}
\label{fig1} 
\end{figure}

\begin{figure} 
\centering
\mbox{\psfig{figure=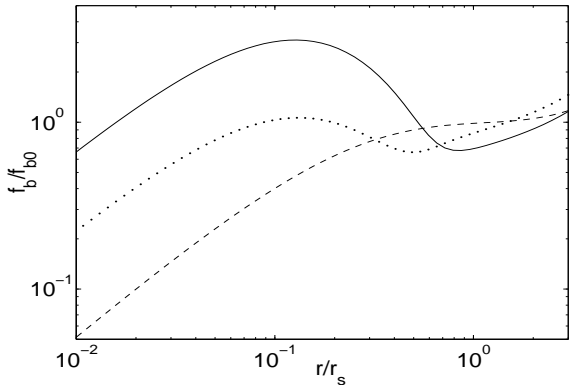,height=2.0in,width=3.0in}} 
\caption{The same as the previous figure, but for the 
baryonic mass fraction (divided by the primordial value).} 
\label{fig2} 
\end{figure}

\begin{figure} 
\centering
\mbox{\psfig{figure=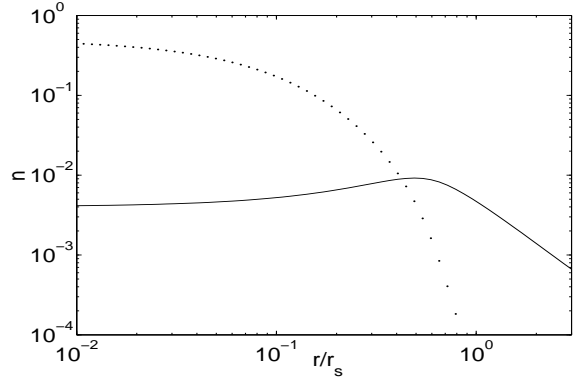,height=2.0in,width=3.0in}}
\caption{The number density of hydrogen (solid) and 
helium (dashed) as a function of radius. 
The curves  are normalized
so that the total number of particles inside $3r_s$ is unity.}
\label{fig3}
\end{figure}

The constants $C_1$ and $C_2$ in the analytic solution
(\ref{analytic:sol}) are fixed by the boundary conditions imposed on
the abundances.  Here we require that the ratio of total helium to
hydrogen abundances inside the virial radius is equal to the
primordial value ($\sim 0.08$).  We will present results for a cluster
with gas temperature such that $\eta=10$ and a virial radius equal to
$3r_{\rm s}$ in agreement with observations and N-body simulations of
massive clusters (NFW 1997; Ettori \& Fabian 1999).  In
Fig. \ref{fig1} we show as the solid line the baryonic density
obtained from the analytic solution.  For comparison we also plot as
the dashed curve the profile (\ref{nodiff}) which corresponds to
equilibrium without diffusion. A proper estimation of the density
profile from the observed X-ray emissivity ($\propto (\sum n_{\rm i}
Z_{\rm i}^2 \sum n_{\rm i} Z_{\rm i})$) should take into account
variations of the He/H abundance ratio throughout the ICM.  Assuming
constant He/H ratio can yield a biased estimate of the profile. To
demonstrate this we plot as the dotted line in fig.~\ref{fig1} the
estimated profile if a constant abundance ratio were assumed.  In
fig.~\ref{fig2} we show the baryonic mass fraction as a function of
radius for the same three cases as in fig.~\ref{fig1}.  We see (solid
line) that diffusion introduces distinct features in the behavior of
the baryonic fraction as a function of radius.  Finally
fig. \ref{fig3} shows the number density of helium and hydrogen in the
case with diffusion.

\section{Conclusions}

We have seen that diffusion steepens the gas density
profile near the center of hot clusters, increasing their X-ray
luminosity (by factor of 5 in our example).  In colder clusters
($T\sim 10^7$ K) the diffusion time-scale is larger than the Hubble
time and their luminosity remains unchanged. This may explain the
observed discrepancy between the observed $L\propto T^{3}$ (Mushotzky
1984; Edge \& Stuwart 1991; David et al. 1993) and $L\propto T^{2}$
which is expected from self-similar arguments (e.g., Kaiser 1986).

If the inner regions of clusters are dominated by helium then the
baryonic mass density as inferred from the X-ray emissivity can be
underestimated by $\sim 30\%$ if constant He/H abundance ratio is
assumed, while in the helium-deficient outer region it can be
overestimated by $\sim 7\%$.  Estimates of of the dark matter density
can be affected by even a larger factor. These estimates 
assume hydrostatic equilibrium (eq. \ref{hstat}), 
so by taking $\mu= 0.59$ (cosmic abundance)
instead of 0.5 (pure hydrogen plasma) in the outer regions, we
underestimate the total mass by $\sim 18
\%$ (Qin \& Wu 2000).  In the helium dominated 
 core the mass would be overestimated by a factor of $2.3$.

 Our
  estimates of the sedimentation time-scales have neglected magnetic
  fields.  Magnetic fields with coherence length comparable to 
  the size  of the cluster force the ions to move on longer orbits defined
  by the field lines.  This can increase the sedimentation time-scales
  by factor of a few.  Small-scale
  magnetic fields, however,    can increase the time-scales by a factor ranging
  from a few in some estimates (Narayan \& Medvedev 2001; Malyshkin
  2001) to 100-1000 in others (Chandran \& Cowley 1998).

\section*{acknowledgment}
We thank Marat Gilfanov for stimulating discussions.
This
research was supported by the Technion V.P.R Fund- 
Henri Gutwirth Promotion of Research Fund,  the German-Israeli Foundation
for Scientific Research and Development, and the EC RTN network 'The physics of the intergalactic matter'.

\end{document}